# MULTI-BUNCH DYNAMICS IN RF PHOTOINJECTORS THROUGH AN ENVELOPE EQUATION APPROACH

M.Ferrario, INFN-LNF, Frascati, Italy
L.Serafini, INFN, Milano, Italy

*Abstract*

Free Electron Lasers driven by Super-Conducting Linacs require the generation, acceleration and transport of trains of bunches with high charge, high repetition rate and low emittance. A numerical model already developed for the modeling of beam dynamics in SC Linacs has been now extended to treat RF guns with proper description of the photocathode bunch generation. The model algorithm is based on a coupled integration of Newton and Maxwell equations under a slowly varying envelope approximation for the time evolution of mode amplitudes in the accelerating structure and an envelope equation description for a set of bunch slices. In this way beam loading effects and higher order modes excitation can be studied. The application to the TTF-FEL injector as a particular design is presented and the optimization according to the invariant envelope concept is discussed, in the frame of single bunch calculations compared to the results of other multi-particle codes.

## 1 INTRODUCTION

Typical beams delivered by SC Linacs at intermediate energies (hundreds of MeV) with quality consistent to FEL requirements (very low emittances at high peak current) are affected by multi-bunch effects in the Linac (due to the high repetition rate in the train as well as the long filling times of the cavities) and, at the same time, by space charge effects, which are still relevant at these energies to cause emittance degradation due to correlation. Single bunch simulations of the beam dynamics performed from the photo-cathode surface up to the undulator injection at the Linac exit are at the far limit of multi-particle codes like PARMELA, because CPU times of several hours are typically required. Due to the non-linearity of the problem, based on space charge correlations within the bunch, standard rms linear space charge description (as performed by matrix-based transport codes like TRACE3D) are not able to describe the associated emittance oscillations, as extensively discussed in Ref. [1].

When multi-bunch effects are to be taken into account, because of the high repetition rate in the train and, as well, because of fluctuations in the pulse intensity and timing jitters of the laser driving the photo-cathode, a multi-particle simulation would be unaffordable.

For this reason we recently enhanced the capability of the code HOMDYN [2], developed originally to describe multi-bunch effects in Linacs under a fast running envelope description, to model the beam generation at the photo-cathode in a photo-injector RF cavity. Higher mode excitation below cut-off as well as non-relativistic beam dynamics were already being modeled by the code, as reported in Ref. [2].

In Section 2 we describe the formalism based on envelope equation applied to a family of representative bunch slices, used by the code to predict the emittance behavior in the photo-injector and in the Linac, while in Section 3 we show a comparison with a multi-particle code, anticipating a good agreement on beam emittance and energy spread, but with one order of magnitude faster CPU time. A full simulation from the cathode up to the second cryomodules exit of the TTF SC Linac at 230 MeV, is reported in the last Section. Emittance oscillations due to cold-plasma like oscillations in the beam envelope, still within the laminar (*i.e.* space charge dominated) regime at this energy, have been found in quite good agreement with the prediction of the theory reported in Ref.[1].

The analytical study developed there is based on a similar approach than the one followed in HOMDYN, *i.e.* the beam dynamics and associated emittance damped oscillations are explained as plasma oscillations proceeding at the same frequency for all the set of slices in which the bunch can be represented. The optimum operating condition assuring the minimum emittance at the transition from the laminar to the thermal regime has been predicted to be the acceleration of the beam under the Invariant Envelope condition, a particular equilibrium mode for an accelerated beam analogous to Brillouin flow in drifting beams: one of the implications of such a beam transport mode is the need to achieve a careful matching of the beam at the exit of each accelerating section as far as the beam runs in the laminar regime, so that the beam envelope oscillations around the corresponding Brillouin flow are as small as possible.

The laminar regime extends up to an unexpected high energy level given by:

$$\gamma = \sqrt{\frac{8}{3} \frac{\hat{I}}{2 I_o \varepsilon_{th} \gamma'}} \qquad (1)$$

where $\hat{I}$ is the peak current, $I_O$=17 KA the Alfven current, $\gamma' = \dfrac{e E_{acc}}{m_e c^2}$ and $\varepsilon_{th}$ the thermal emittance (typically 0.5 mm mrad). For a TTF-FEL beam [3] where $\hat{I}$=60 A and $E_{acc}$=15 MV/m the transition occurs at

$\gamma=272$, implying that most of the TTF Linac is sensitive to possible emittance growths of this nature.

In order to achieve the matching, the accelerating gradient at the entrance of accelerating modules and the focusing gradient of the intermediate focusing elements (solenoids) must respect the following rules:

$$\gamma' = \frac{4}{R}\sqrt{\frac{\hat{I}}{3I_o\gamma}} \quad \text{and} \quad B_z = \frac{2}{R}\sqrt{\frac{2\hat{I}}{I_o\gamma}}\frac{m_e c}{e} \quad (2)$$

With the additional constrain R'=0 (parallel envelope). These predictions are shown to be very well satisfied by the simulations presented in the last Section.

## 2 NUMERICAL MODEL

The basic approximation in the description of beam dynamics lays in the assumption that each bunch is described by a uniform charged cylinder, whose length and radius can vary under a self-similar time evolution, keeping anyway uniform the charge distribution inside the bunch. By slicing the bunch in an array of cylinders (Multi-Slices Approximation), each one subject to the local fields, one obtains also the energy spread and the emittance degradation due to phase correlation of RF and space charge effects.

The radial space charge fields (linear component) at a distance $\zeta_s = z_s - z_t$ of the $s^{th}$ slice from the bunch tail located at $z_t$, is given by:

$$E_r^{sc}(\zeta_s, A_{r,s}) = \frac{Q}{4\pi\varepsilon_o R_s L}G(\zeta_s, A_{r,s})$$

$$G = \frac{\zeta_s/L}{\sqrt{(\zeta_s/L)^2 + A_{r,s}^2}} + \frac{1-\zeta_s/L}{\sqrt{(1-\zeta_s/L)^2 + A_{r,s}^2}}$$

where $Q$ is the bunch charge, $L$ the bunch length, $R_s$ the slice radius and $A_{r,s} \equiv R_s/(\gamma_s L)$ is the slice rest frame aspect ratio.

The evolution of each slice radius $R_s$ is described into the time-domain according to an envelope equation, including thermal emittance (first term), solenoid focusing field (second), space charge effects (third), image charges from the cathode surface (fourth) RF-focusing (fifth) and a damping due to acceleration (sixth):

$$\ddot{R}_s = \left(\frac{4\varepsilon_n^{th} c}{\gamma_s}\right)^2 \frac{1}{R_s^3} - \left(\frac{eB_z(z_s)}{2m_o\gamma_s}\right)^2 R_s +$$

$$\frac{2c^2 k_p}{R_s \beta_s \gamma_s^3}G(\zeta_s, A_r) - \frac{2c^2 k_p}{R_s \beta_s \gamma_s}(1+\beta_s^2)G(\xi_s, A_r) +$$

$$-K_s^{rf} R_s - \beta_s\gamma_s^2\dot{\beta}_s\dot{R}_s$$

where the dots indicate the derivation with respect to time, $\varepsilon_n^{th}$ is the rms normalized thermal beam emittance, $k_p = \frac{I}{2I_o}$ is the beam perveance and the RF focusing gradient $K_s^{rf} = \frac{e}{2\gamma m_o}\left(\frac{\partial E_z}{\partial z} + \frac{\beta_s}{c}\frac{\partial E_z}{\partial t}\right)$ has been expressed through the linear expansion off-axis of the accelerating field $E_z = E_z(0,z,t)$.

We use the following expression for the correlated rms emittance: $\varepsilon_n^{cor} = \frac{1}{2}\sqrt{\langle R^2 \rangle \langle (\beta\gamma R')^2 \rangle - \langle R\beta\gamma R' \rangle^2}$ where $R' = \frac{d}{dz}R$ and the average $\langle \ \rangle = \frac{1}{N}\sum_{s=1}^{N}$ is performed over the N slices. The total rms emittance will be given by a quadratic summation of the thermal emittance and the correlated emittance: $\varepsilon_n = \sqrt{\left(\varepsilon_n^{th}\right)^2 + \left(\varepsilon_n^{cor}\right)^2}$.

An effective description of the bunch generation has been recently added to the code. For a given laser pulse time length $\Lambda$ a neutral bunch ($\tilde{Q}=0$) of initial energy of some eV and initial length $L_0 = \beta c \Lambda$ is generated behind the cathode location $z_C$, where the accelerating field is zero, with the barycenter located at $z_b = z_c - L_0/2$. At t=0 the bunch starts moving according to the equation of motion and slice by slice enters in to the RF gun cavity where each slice is activated in the envelope equation sharing an increasing bunch charge: $Q_{act} = Q \cdot \left(1 - \frac{z_c - z_t}{L_o}\right)$

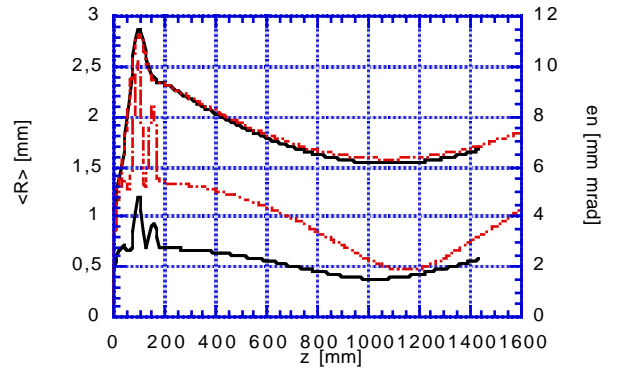

Figure 1: Envelope and rms normalized emittance and as computed by ITACA (solid line) and HOMDYN (dashed line). Thermal emittance set to zero.

A cross check of the HOMDYN model has been performed with respect to a completely different simulation tools: ITACA [4], a C.I.C. self consistent electromagnetic code. The agreement between the two codes is so remarkable, as shown in Fig 1, that clearly proofs the validity of an envelope equation based description for this specific beam physics case. The RF gun taken into consideration is a 1+1/2 cell operated at 1.3 GHz, producing a 1. nC bunch charge at 80 A peak current: the peak RF field at the cathode is 50 MV/m with as laser spot size (hard edge) of 1.5 mm. Applying a 1.8 kG peak solenoid field, the beam envelope obtained is plotted in Fig.1 versus the beam propagation distance z from the cathode surface (z=0) down through the gun (175 mm long) and the following drift section. It is clearly visible the strongly space charge dominated behavior of the envelope, going through a gentle laminar waist in the drift. The rms normalized transverse emittance is also plotted, clearly showing the emittance correction process.

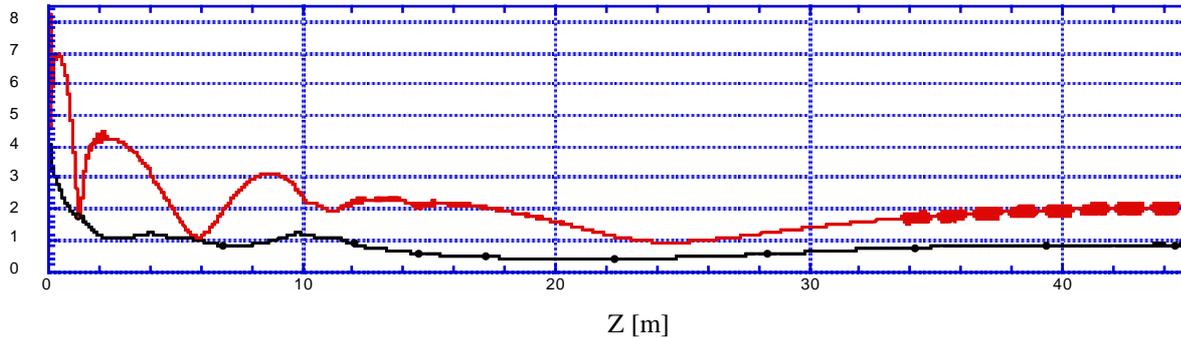

Figure 2: Beam envelope [mm] (line with dots) and rms normalized emittance [mm mrad] (plain line) up to 230 MeV.
Beam line: Gun [0<z<0.18], Drift [0.18<z<1.2], Booster [1.2<z<2.4], Drift [2.4<z<11.4],
Cryomodule1 [11.4<z<22.6], Drift [22.6<z<33.6], Cryomodule2 [33.6<z<44.8].

## 3 TTF-FEL LINAC CASE

We report in this section an analysis of the TTF-FEL [3] beam envelope and emittance oscillations up to the end of the second cryomodule, as shown in Fig 2, corresponding to a beam energy of 230 MeV. The 1 nC beam is generated in one and half cells RF-gun operating at 1.3 GHz with 50 MV/m peak field on the cathode, by a 12 ps long laser pulse with 2.5 mm radius. The gun is embedded in a 0.193 T solenoid field. In the drifting tube downstream the gun the rms emittance reach a first minimum (1.84 mm mrad) corresponding to the minimum (R'=0) envelope (1.79 mm) at z=1.2 m from the cathode, where the booster is placed. To match the booster cavity (9 cells 1.3 GHz superconducting cavity) to the beam we use eq (2) to compute the accelerating field needed. It results to be 12 MV/m with I=62 A and $\gamma$ = 11. Downstream the booster the rms emittance reach a second minimum (1.108 mm mrad) in the long drift section foreseen to house focusing and diagnostic elements and a chicane compressor. It has been also proposed [5] to use a long solenoid around the beam line to limit the envelope and emittance oscillation around the minimum. The same result can be achieved also by splitting the solenoid in three shorter lenses, providing a small envelope oscillation around the envelope equilibrium. Indeed, at the booster exit (z=2.4 m) the beam envelope (1.04 mm) is still parallel to the z-axis with $\gamma$ = 38 and it would require a 0.043 T long solenoid. We allowed the envelope to diverge up to 1.227 mm where the matching long solenoid (L=8 m) would be 0.0366 T and we splitted it in three lens whose length is l=0.16 m with an equivalent field $B_l = B_L \sqrt{\dfrac{L}{l}} = 0.29$ T for the first and third and 0.23 for the middle one, so that to recover a parallel envelope of 1.04 mm at the entrance of the first cryomodule (z=11.4 m) with 12 MV/m accelerating field. Envelope and emittance oscillations are damped during acceleration but still two emittance oscillations take place inside the first cryomodule with an absolute emittance minimum (0.94 mm mrad) in the downstream drift.

As predicted by the theory the lack of control of the envelope in the drift causes another emittance oscillation in the second cryomodule, that would be avoide by properly positioning focusing elements

## 4 CONCLUSION

The new version of the code HOMDYN [2], implemented with an effective description of the beam generation and transport in rf photoinjectors, represents a fast and powerful tool for single and multi-bunch computations needed to design Linacs for the production of bright electron beams.

## REFERENCES

[1] L. Serafini, J. B. Rosenzweig, "Envelope analysis of intense relativistic quasilaminar beams in rf photoinjectors: a theory of emittance compensation", Phys. Rev. E, **55**, (1997)

[2] M. Ferrario et al., "Multi-bunch energy spread induced by beam loading in a standing wave structure", Part. Acc., **52** (1996), or pdf version in http://ares.lnf.infn.it

[3] "A VUV Free Electron Laser at the TESLA Test Facility at DESY, Conceptual Design Report", TESLA-FEL note 95-03

[4] L.Serafini and C.Pagani, Proceedings of I EPAC, Rome, 1988,(ed. World Sci., Singapore) p.866.

[5] L. Serafini, J. B. Rosenzweig, Proceedings of PAC, Vancouver, Canada, 1997